# Assessment of the reliability of Deconvolution Procedures for RCF Spectroscopy of Laser-Driven Ion Beams

S. McCallum[a,b], G. Milluzzo[c,a], M. Borghesi[a], A. Subiel[b], F. Romano[d]

[a] *Centre for Plasma Physics, Queen's University Belfast,*
*BT7 1NN, United Kingdom*

[b] *Medical Radiation Science, National Physical Laboratory,*
*Teddington, TW11 0LW, United Kingdom*

[c] *Istituto Nazionale di Fisica Nucleare, Laboratori Nazionali del Sud,*
*Via S. Sofia 62, 95123 Catania, Italy*

[d] *Istituto Nazionale di Fisica Nucleare, Sezione di Catania,*
*Via S. Sofia 64, 95123 Catania, Italy*

*E-mail*: smccallum05@qub.ac.uk

ABSTRACT: Laser-driven ion beams are defined by a number of unique features, including a large spread in energy. A stack configuration of radiochromic film (RCF) can be utilized to characterize such beams through measurements of their energy spectra. A spectroscopic procedure is reported that allows the proton energy density within each active layer of a radiochromic film (RCF) stack to be retrieved. This is based upon a deconvolution algorithm developed through Geant4 Monte Carlo simulations to correct the contributions of energy depositions within a given film layer. Through Monte Carlo calculations, the spectrum retrieved from a simulated film stack can be retrieved and compared with a known energy spectrum, providing an examination of the efficacy of this tool. Application of the developed deconvolution procedure thus offers the potential to correctly reconstruct the incident energy spectrum of a laser-driven proton and ion beam from a stack of irradiated RCF.

## Contents

## 1. Introduction

Whilst laser-driven proton and light ion acceleration has attracted significant interest for over 20 years [1, 2], conducting accurate measurements of these beams has proven to be technically challenging [3-5]. In particular, the ultra-high dose rates and wide spectral distributions make conventional measurement techniques impracticable [6-8]. For applications, including clinical and radiobiological ones requiring a precise energy selection, characterisation of such beams through accurate measurement of their energy spectra is necessary. Spectroscopic methods reliant on stacked configurations of radiochromic films (RCF) are well-established for measurements of accelerated proton beams, with several approaches of radiochromic film imaging spectroscopy (RIS) reported in the literature [9-14]. A stacked configuration of films placed perpendicularly to the beam orientation can be used to perform an energy resolved measurement of an impinging ion beam. Differential energy loss results in each particle depositing a fraction of its initial kinetic energy on every film it passes before coming to arrest. For polyenergetic sources such as laser-driven beams, a superposition of kinetic energy contributions is amassed across the films, requiring a calculation for correction of higher energies. This is achieved through a deconvolution or unfolding of the energy transferred to each film in the stack, so that only the particles stopping within a given film remain. The aim of the work reported here was to investigate and assess a developed algorithm for spectroscopy of laser-driven proton and ion beams through Monte Carlo simulations, studying the possible limitations. This procedure requires knowledge of the RCF energy sensitivity values, and an algorithm to unfold the proton energy spectrum from the RCF response, both of which have been evaluated using the Geant4 toolkit [15-17]. Further, the same Monte Carlo methods were utilised to conduct analysis of the performance and limitations of the developed technique in acquiring the energy spectrum. Once validated, the spectroscopic procedure reported offers the potential to reliably extract the laser-driven proton spectra from a stack of irradiated RCF.

## 2. Methodology

Energy resolved measurements of impinging proton and ion beams can be performed using multiple RCF arranged into a stack configuration. The differing stopping positions for protons of a given energy within an RCF stack, means each layer can be defined by a unique energy sensitivity. This is chosen to correspond to the energy required to generate a Bragg peak at that given depth, defining the energy of protons that will be referred to as peak region protons. Low energy components stop in the first few layers of the stack, whilst higher energies penetrate further downstream, giving a total energy composition of stopping protons, in addition to the fractional contributions of those exceeding the energy sensitivity of a given film layer. Unfolding the peak energy from the total energy deposited within any RCF can be achieved through the development of a deconvolution procedure for proton spectroscopy. This relies on an algorithm utilising weight factors to describe the fractional contributions of each energy component within every film. This process is detailed in figure 1.

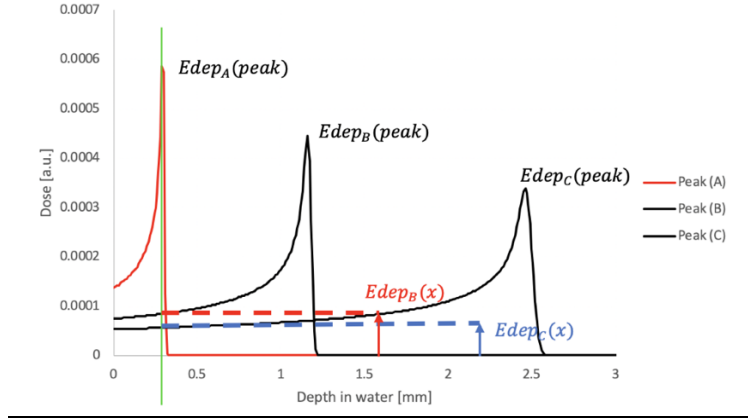

**Figure 1.** Visual representation of the calculation of weight factors. The water equivalent depths of the active layers, in addition to the energy required to produce a Bragg peak at the depth of each, are both well-known. Extrapolating the peak contributions allowed weighting factors to be calculated through normalization of the deposited energy contribution to that of the respective peak value. For example, to calculate the weighting factor provided by peak B to peak A, the ratio of the energy deposited by peak B at the position of peak A, $Edep_B(x)$, to the maximum ionization of B itself, $Edep_B(peak)$, is found. This process is performed for each energy component, at each active layer depth, and a matrix of weight factors is then constructed.

The developed algorithm performs a backwards weighted subtraction of contributions, starting from the final layer, as a singular energy is contained on this film. Careful subtraction of weighted components discriminates the energy of stopping protons within each film from passing energies. This remaining peak or stopping energy is then converted into a measurement of the stopping particle fluence through the corresponding stopping power of every given layer.

$$N_{protons} = \frac{\left|\frac{dE}{dx}\right|^{weighted}_{RCF}}{\left|\frac{dE}{dx.\rho r}\right|_{Geant4}} \quad (Eq. 1),$$

The numerator of Eq. 1 represents the remaining peak stopping energy within every active layer after the deconvolution algorithm has been applied to the total deposited energy within each. The denominator denotes the energy transfer as a function of the thickness of film material crossed, found through Monte Carlo simulation. A processing script was written using the MATLAB software [18], that compiles all of the required input parameters and procedures of this spectroscopic method into a single program. This provides the possibility to directly input scanned

RCF images, and through simple modification, data from simulation, for a direct reconstruction of the proton energy spectrum. A typical reconstructed spectrum is highlighted in figure 2, with data obtained at a laser-driven proton facility.

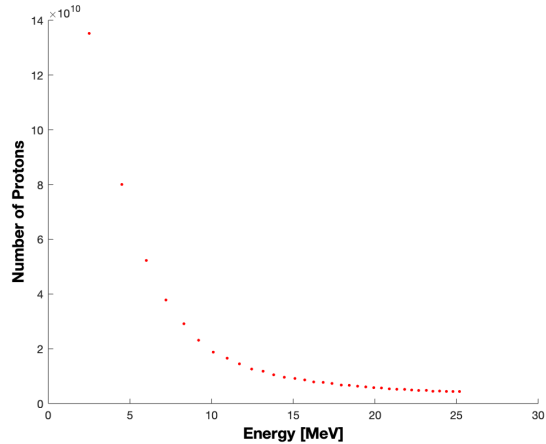

**Figure 2.** Proton energy spectrum found from an irradiated stack of RCF of the model GafChromic HDV2. This data was taken from a laser-plasma experiment at the LULI facility (Laboratoire pour l'Utilisation des Lasers Intenses, École Polytechnique, France). The stack was placed immediately after the target, from which protons were generated with the typical TNSA exponential behaviour.

The resultant energy spectrum displays an exponentially decreasing behaviour typical of laser-driven beams produced through the target normal sheath acceleration process [1, 2]. The fact that this is observed in the data in figure 2 gives some confidence that the procedure can reproduce the expected spectral profile.

## 3. Monte Carlo Analysis

To further assess the effectiveness of the developed deconvolution procedure, a Monte Carlo analysis was conducted through Geant4. A replicated RCF stack configuration of the model GafChromic EBT3, with symmetrical structure of a 28 $\mu m$ active layer, sandwiched between two polyester dead layers, was constructed as outlined within the manufacturer's specifications [19]. Detailed simulation of this film stack is a vital first step in the spectroscopic procedure development, allowing the energy sensitivity and corresponding stopping power values to be evaluated for each film layer, in addition to the weighting factors required in the deconvolution algorithm.

The reliability of the developed deconvolution procedure as a tool for spectroscopy was assessed through examination of the retrieved deconvolution spectrum, with one that is known. Through Geant4 simulation, a proton source with tailored energy could be sent into the constructed RCF stack, and the deposited energy converted to a measurement of the particle number (energy fluence) at each active layer node using the deconvolution algorithm developed. This arrangement was used for input proton sources with both exponential and flat energy spectra. The latter of these source spectra proved more useful in highlighting potential discrepancies between the actual and expected spectra. From analysis, it was noticed that particularly for laser-driven energy spectra, with particle numbers extending orders of magnitude, the differences between the retrieved spectral particle numbers can be quite large, whilst still maintaining an apparently good degree of agreement. During cross-comparison, the potential to disguise discrepancies between spectra was reduced with the use of a flat spectrum. Once the proton energy spectrum had been recovered from the energy deconvolution data, it was cross-compared with the original energy spectrum. A measurement of the spectrum that originates at the source can be obtained from the simulation through examination of the particle flux at a thin region coinciding with the front face of the film stack. This eliminates interaction of the impinging proton beam with the RCF material, and potential errors induced through conversion of the measured deposited energy to particle flux. Analysis of the retrieved spectrum through application of the deconvolution algorithm in Geant4 is shown in figure 3.

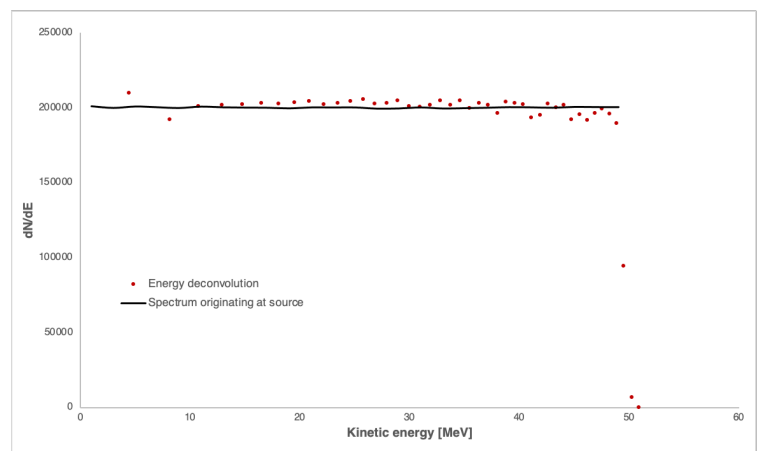

**Figure 3.** Cross-comparison of the proton energy spectrum obtained through a deconvolution of the total energy deposited in each film layer, with the proton fluence spectrum originating at the source as measured at the stack entrance.

A reasonable agreement between the deconvolution and entrance spectra is observed from fig. 4, outlining the accuracy of the developed procedure in obtaining the correct particle flux at each measurement node. This systematic Monte Carlo investigation thus gives an insight into the working order of the algorithm for deconvolution, providing an indication of its reliability in correctly reconstructing the energy spectrum. Within previous works concerning RCF spectroscopy, the final spectrum is often assumed to be correct, with no such systematic check performed. Analysis has shown that this cannot be taken for granted, and so by carrying out this procedure some confidence is gained concerning the reliability of this spectroscopic tool.

## Conclusions

A spectroscopic procedure for the measurement of laser-driven proton energy spectra based on the use of a stacked configuration of radiochromic films has been developed and reported here. A deconvolution algorithm that operates through an iterative backwards weighted subtraction of energy components from successive films has been developed to unfold the stopping proton energy from the total energy deposited in each film layer. Initial tests demonstrated reconstruction of a typical exponential-like spectrum with large energy spread for films irradiated using a laser-driven proton beam. Further analysis of the developed spectroscopic procedure was conducted through Monte Carlo methods utilising the Geant4 particle simulation toolkit. Comparison of the spectrum retrieved through deconvolution of the energy transferred to each film, to that originating at the source for a flat energy spectrum showed a good agreement, indicating the applicability of this tool in the spectral reconstruction of a laser-driven proton source. Although the analysis reported is promising, a thorough examination of experimental data should be carried out to validate the developed procedure. A reasonable result would outline the potential of this tool in deriving a fast measurement of the energy spectrum from an irradiated stack of radiochromic films. Nonetheless, this systematic investigation based on analysis of spectral deconvolution through detailed Monte Carlo simulations represents one that has not been tried before. Through cross-comparison within simulation, this has allowed an effective evaluation of the performance of such a spectroscopic tool required for accurate measurement of the proton energy spectrum generated through laser-driven beams.